\documentclass[conference]{IEEEtran}
\IEEEoverridecommandlockouts
\usepackage{cite}
\usepackage{amsmath,amssymb,amsfonts}
\usepackage{algorithmic}
\usepackage{graphicx}
\usepackage{textcomp}
\usepackage{xcolor}
\def\BibTeX{{\rm B\kern-.05em{\sc i\kern-.025em b}\kern-.08em
    T\kern-.1667em\lower.7ex\hbox{E}\kern-.125emX}}
\begin{document}

\title{Does EDPVR Represent Myocardial Tissue Stiffness? Toward a Better Definition}

\author{\IEEEauthorblockN{Rana Raza Mehdi}
\IEEEauthorblockA{\textit{Department of Biomedical Engineering} \\
\textit{Texas A\&M University}\\
College Station, TX, USA \\
Email: razamehdi@tamu.edu}

\and
\IEEEauthorblockN{Emilio A. Mendiola}
\IEEEauthorblockA{\textit{Department of Biomedical Engineering} \\
\textit{Texas A\&M University}\\
College Station, TX, USA \\
Email: emilio.mendiola@tamu.edu}
\and
\IEEEauthorblockN{Vahid Naeini}
\IEEEauthorblockA{\textit{Department of Biomedical Engineering} \\
\textit{Texas A\&M University}\\
College Station, TX, USA \\
Email: vahid\_naeini@tamu.edu}
\and
\IEEEauthorblockN{Gaurav Choudhary}
\IEEEauthorblockA{\textit{Division of Cardiology} \\
\textit{Warren Alpert Medical School of Brown University}\\
Providence, RI, USA \\
Email: gaurav\_choudhary@brown.edu}
\and
\and
\IEEEauthorblockN{Reza Avazmohammadi}
\IEEEauthorblockA{\textit{Department of Biomedical Engineering} \\
\textit{Texas A\&M University}\\
College Station, TX, USA \\
Email: rezaavaz@tamu.edu}

}

\maketitle

\begin{abstract}
Accurate assessment of myocardial tissue stiffness is pivotal for the diagnosis and prognosis of heart diseases. Left ventricular diastolic stiffness ($\beta$) obtained from the end-diastolic pressure-volume relationship (EDPVR) has conventionally been utilized as a representative metric of myocardial stiffness. The EDPVR can be employed to estimate the intrinsic stiffness of myocardial tissues through image-based in-silico inverse optimization. However, whether $\beta$, as an \textit{organ-level} metric, accurately represents the \textit{tissue-level} myocardial tissue stiffness in healthy and diseased myocardium remains elusive. We developed a modeling-based approach utilizing a two-parameter material model for the myocardium (denoted by $a_f$ and $b_f$) in image-based in-silico biventricular heart models to generate EDPVRs for different material parameters. Our results indicated a variable relationship between $\beta$ and the material parameters depending on the range of the parameters. Interestingly, $\beta$ showed a very low sensitivity to $a_f$, once averaged across several LV geometries, and even a negative correlation with $a_f$ for small values of $a_f$. These findings call for a critical assessment of the reliability and confoundedness of EDPVR-derived metrics to represent tissue-level myocardial stiffness. Our results also underscore the necessity to explore image-based in-silico frameworks, promising to provide a high-fidelity and potentially non-invasive assessment of myocardial stiffness.
\end{abstract}


\section{Introduction}
The end-diastolic pressure-volume relationship (EDPVR) serves as a critical metric, offering essential insights into the diastolic function of cardiac physiology \cite{Numata}. The EDPVR is considered the gold-standard measurement of chamber compliance and is used by clinicians and researchers to characterize altered LV chamber mechanical behavior in diastole \cite{Bermejo2013}. However, myocardium behavior is inherently complex, and conditions such as fibrosis or altered myofiber orientation can significantly alter myocardial stiffness without delectably manifesting in EDPVR \cite{Mehdi, Babaei}. Material model parameters integral to in-silico models are typically derived from EDPVR; however, the capacity of the EDPVR to capture the individual subtle influence of those parameters on LV behavior remains understudied. The limitation of the EDPVR to accurately characterize the complex anisotropic behavior of the myocardium will potentially obscure true dysfunction, leading to misdiagnosis or inadequate treatment.

In this work, we investigate the capacity of the EDPVR to describe altered myocardial behavior. For this purpose, we used previously developed rodent computational cardiac models \cite{Avaz2019}. 
These models were used to conduct a parametric study to improve our understanding of how well EDPVR represents altered myocardial mechanical behavior. 

\section{Methodology}
\subsection{Animal model}
A cohort of 13 male rats was used, including Wistar Kyoto, Sprague-Dawley, and Fischer-344 strains. The cohort included healthy rats (n=5) and rats with pulmonary hypertension (PH, n=8). PH was induced as previously described \cite{Mendiola}. 

\subsection{Image reconstruction}
\begin{figure}[htbp]
\centerline{\includegraphics[width=0.5\textwidth]{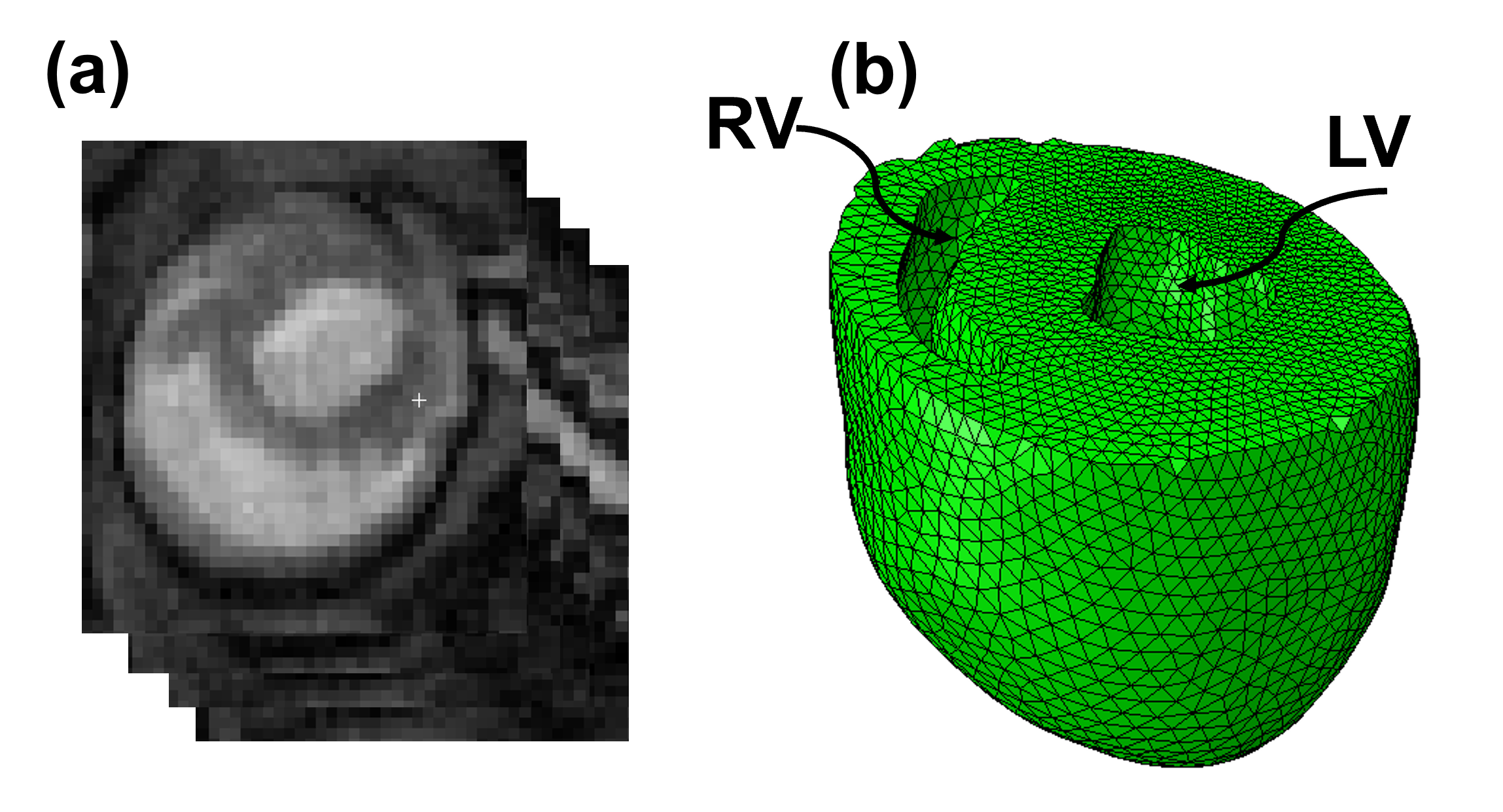}}
\caption{(a) Cardiac magnetic resonance (CMR) scans of rat hearts at end-diastole (ED) were used to reconstruct the 3-D cardiac geometry and develop (b) subject-specific finite-element heart models.}
\label{fig}
\label{fig:Geometry}
\end{figure}

After sacrifice, detailed anatomical imaging of the heart was obtained via high-resolution cardiac magnetic resonance (CMR) scans with the heart at the end-diastolic (ED) configuration, employing a Bruker Biospec 7T scanner. The imaging protocol utilized an isotropic 60 $\mu$m resolution, with a field of view (FOV) of 15 x 15 x 20 mm, echo time ($T_E$) 7.5 ms, and repetition time of ($T_R$) 40 ms. Subsequently, the CMR scans were segmented, and a three-dimensional (3-D) reconstruction of the cardiac geometry of the heart (truncated below the valve plane) was obtained using Materialise Innovation Suite (Materialise, Belgium). The high-fidelity 3-D reconstruction was meshed using quadratic tetrahedral elements to develop a cardiac finite-element (FE) model (Fig. \ref{fig:Geometry}). Detailed methods regarding image acquisition and model development are described in Avazmohammadi et al. \cite{Avaz2019}.

\subsection{Myocardium constitutive model}
The myocardium was modeled as a nonlinear anisotropic hyperelastic material \cite{Avazmohammadi}. In our cardiac simulations, we employed a simplified version of the Holzapfel-Ogden constitutive model \cite{Holzapfel} to represent the mechanical behavior of the myocardium. The strain energy function was expressed as

\begin{equation}
    \Psi = \frac{a}{2b}\{exp[b(I_1-3)]-1\}+\frac{a_f}{2b_f}\{exp[b_f(I_{4f}-1)^2]-1\},       
    \label{HO}
\end{equation}

\noindent where, $a$, $b$, $a_f$, and $b_f$ were material constants, with dimensions of stress for $a$ and $a_f$, while $b$ and $b_f$ were dimensionless. The deformation in the ground matrix and along the fiber direction was captured by $I_1=tr{\bf C}$ and $I_{4f}={\bf f}_0{\cdot}({\bf C} {\bf f}_0)$, respectively, where ${\bf C}$ denoted the right Cauchy-Green deformation tensor.

\subsection{Pressure volume analysis during diastole}

\begin{figure}[htbp]
\centerline{\includegraphics[width=0.5\textwidth]{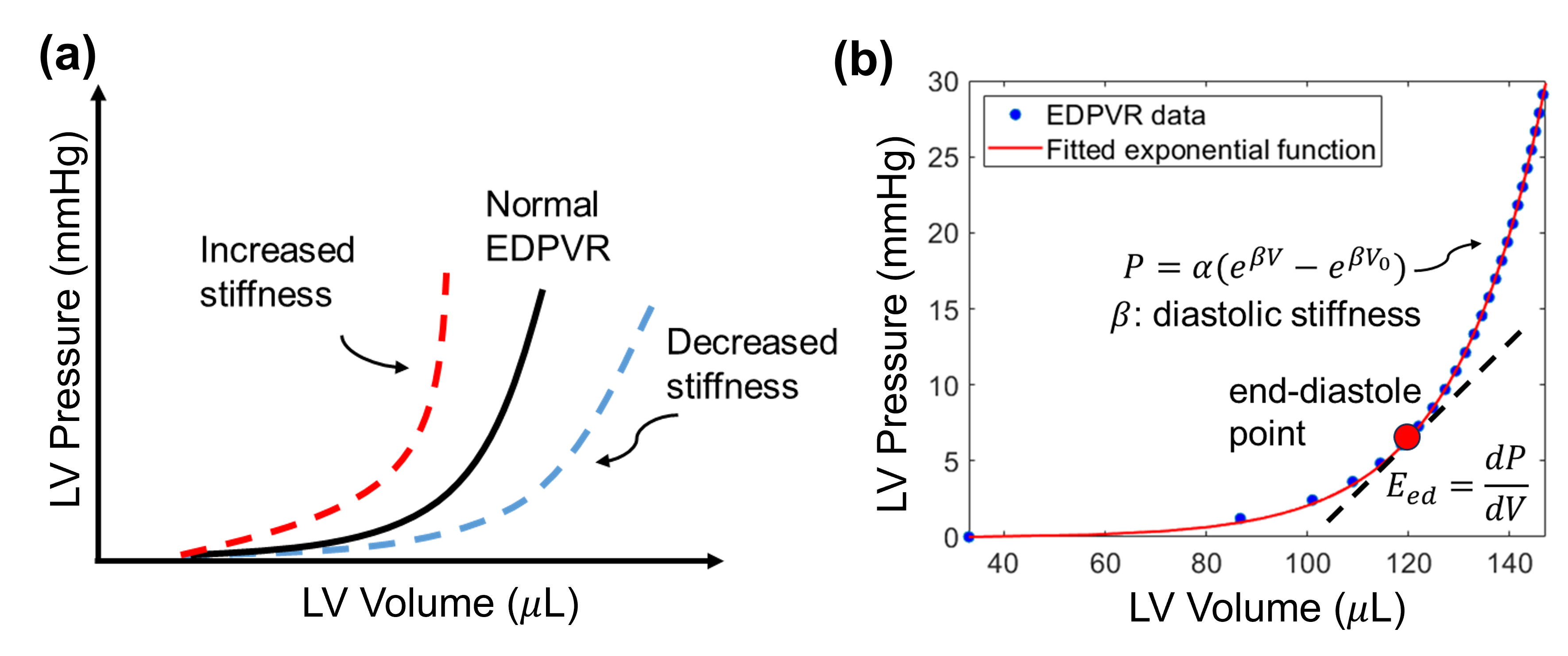}}
\caption{(a) The end-diastolic pressure-volume relationship (EDPVR) qualitative changes due to altered myocardial biomechanical conditions. (b) EDPVR-derived stiffness metric $\beta$ was estimated by fitting the exponential function to the EDPVR. End-diastolic elastance ($E_{ed}$) is the slope of the exponential function curve at end-diastole. dP/dV, change in pressure per change in volume; $V$ and $V_0$ indicate measured volume and volume at zero pressure, respectively.}
\label{fig:schematic}
\end{figure}

The simulated EDPVR for the cardiac model was obtained through forward FE simulations using ABAQUS (Simulia, Providence, RI, USA). The basal surface was fixed in the apex-to-base direction ($Z$), and a linearly ramped pressure was applied to the LV endocardial surface with a maximum pressure of 30 mmHg. The load was applied with 100 equal loading steps. The EDPVR was calculated as alterations in material model parameters $a_f$, and $b_f$ were systematically introduced. This procedure resulted in a set of EDPVR curves corresponding to a fixed cardiac geometry with varying material parameter configurations representing diverse diastolic behaviors in terms of increased and decreased diastolic stiffness (Fig. \ref{fig:schematic}a). Subsequently, an EDPVR-derived metric of diastolic stiffness was obtained from the simulations by fitting the exponential function $P = \alpha (e^{\beta V}-e^{\beta V_{0}})$ to the simulated EDPVR (Fig. \ref{fig:schematic}b) \cite{Jang}. In this equation, $\beta$ represents the myocardial diastolic stiffness, and $V_0$ is the pressure estimated at $P=0$. An analysis of the changes in the EDPVR-derived stiffness metric $\beta$ resulting from the prescribed alterations to the material parameters, $a_f$ and $b_f$, was conducted to discern the nuanced relationship between intrinsic myocardial stiffness and the EDPVR and $\beta$.

\section{Results}
\subsection{Effect of altered myocardial stiffness on EDPVR}
\begin{figure}[htbp]
\centerline{\includegraphics[width=0.5\textwidth]{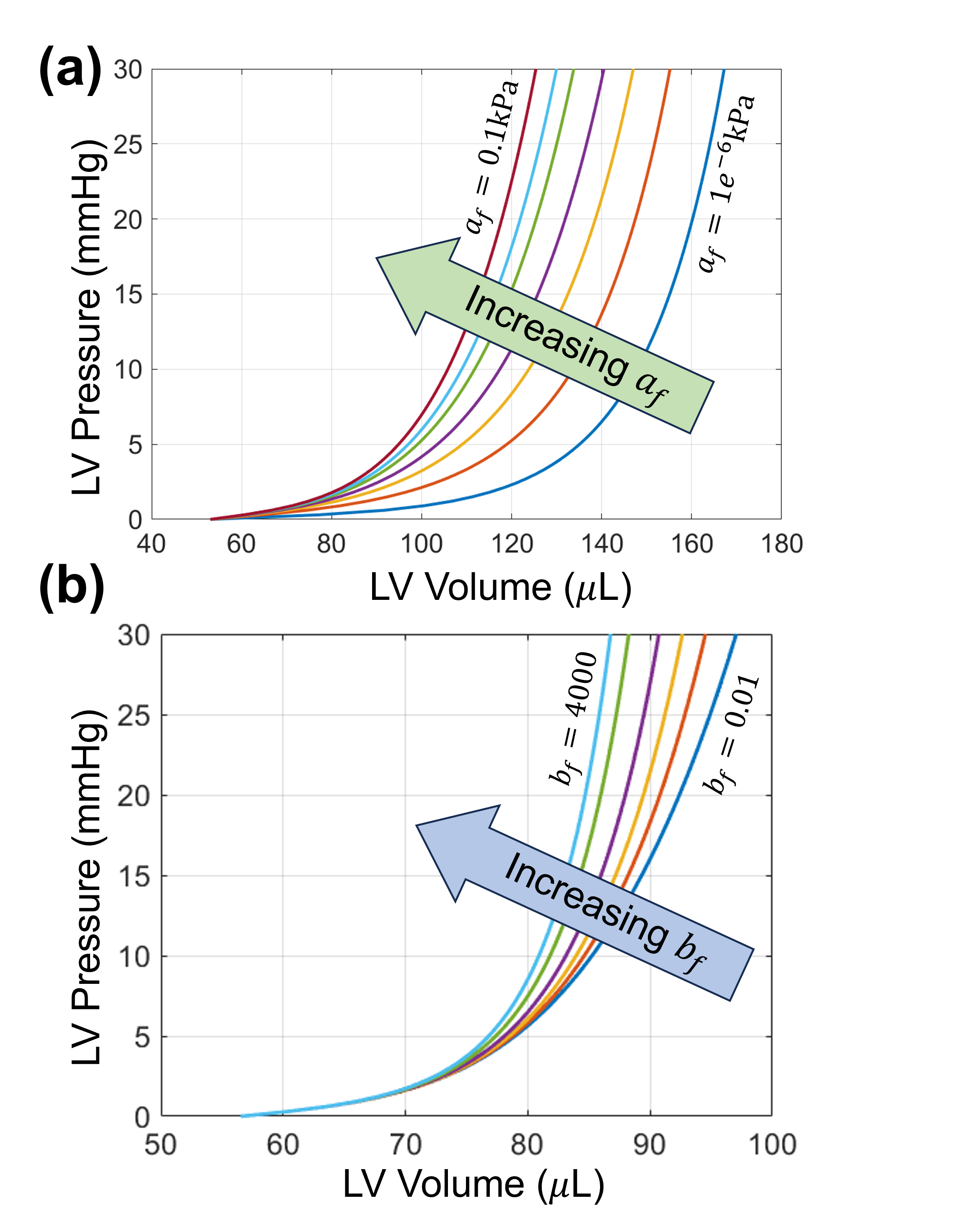}}
\caption{End diastolic pressure-volume relationship (EDPVR) curves shifted leftward when increasing (a) $a_f$ and (b) $b_f$.}
\label{fig:EDPVR}
\end{figure}

Increasing the material parameters, thus effectively increasing myocardial stiffness, consistently induced a leftward shift of the EDPVR curve (Fig. \ref{fig:EDPVR}). Traditional EDPVR analysis would indeed take this leftward displacement to suggest increased myocardial stiffness. The initial increases, starting from low values of $a_f$ and $b_f$, exhibited a more pronounced leftward shift in the EDPVR, emphasizing the substantial influence on the biomechanics of the myocardium. However, the leftward shift became more subtle as the values of the parameters increased further from the initial value suggesting a saturation point where further increases have diminishing effects on the EDPVR.

\subsection{Relationship between material parameters and EDPVR-derived stiffness}
\begin{figure}[htbp]
\centerline{\includegraphics[width=0.5\textwidth]{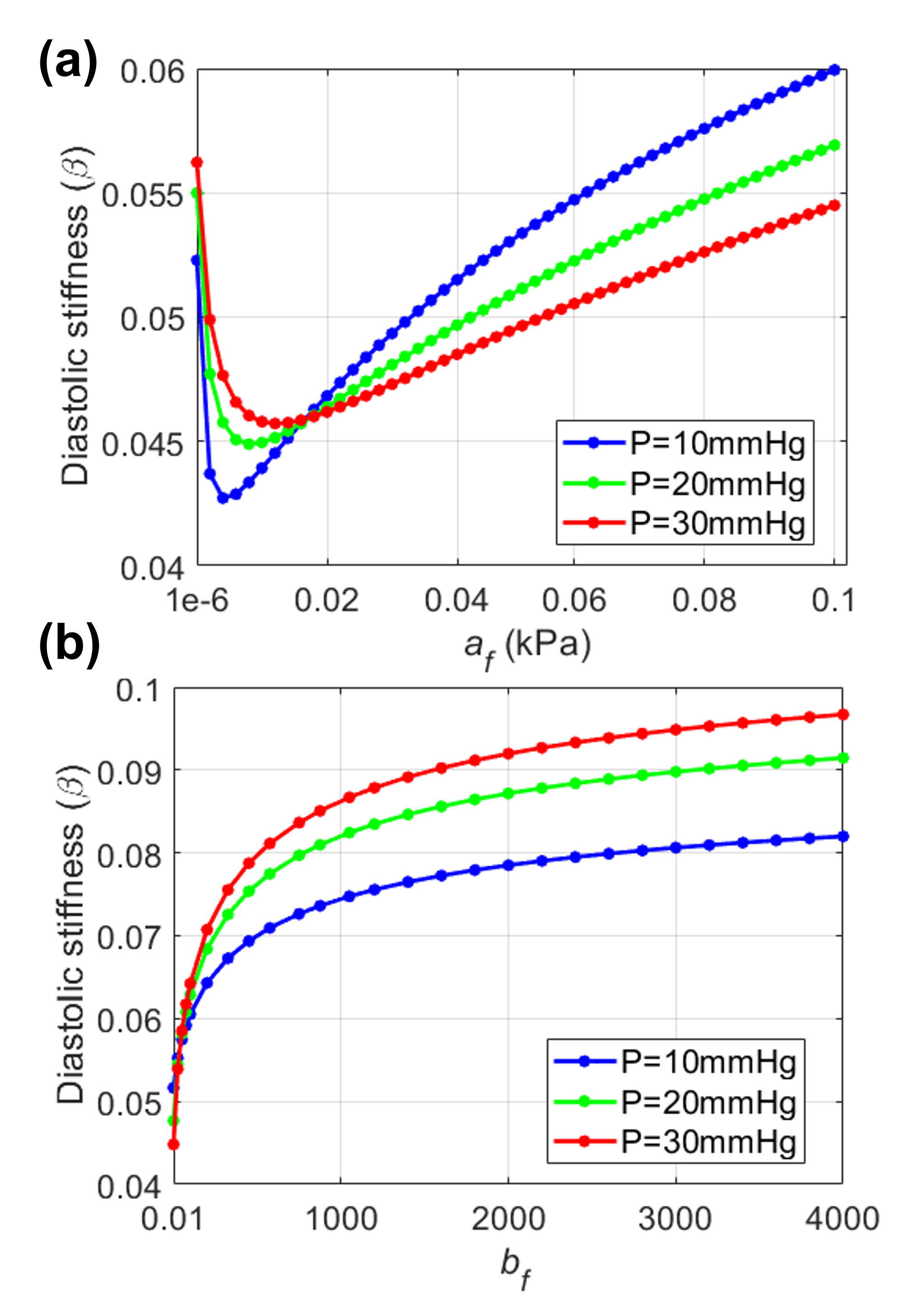}}
\caption{Relationship between EDPVR-derived stiffness metric $\beta$ and myocardial material parameters under various load conditions, including pressures 10, 20, and 30 mmHg for (a) $a_f$ ($b_f$=10) and (b) $b_f$ ($a_f$=0.05kPa).}
\label{fig:sensitivity}
\end{figure}

An analysis of the relationship between myocardial stiffness, characterized by $a_f$ and $b_f$, and EDPVR-defined stiffness provided valuable insights into the dynamic response of $\beta$ to variations in myocardial stiffness. Interestingly, initial increases of $a_f$ resulted in a reduction in EDPVR-derived stiffness, represented by a downward trend in $\beta$ (Fig. \ref{fig:sensitivity}a). However, beyond $\sim a_f$=0.01 kPa, $\beta$ began to increase. This inflection point marks a transition where increases in $a_f$ lead to increased $\beta$.

The relationship analysis between $b_f$ and $\beta$ was distinct from that noted between $a_f$ and $\beta$. Initial increases in $b_f$ resulted in a sharp increase in the value of $\beta$ (Fig. \ref{fig:sensitivity}b). This initial response suggests $b_f$ has a more direct impact on variations in the EDPVR than does $a_f$. However, unlike the trend observed with $a_f$, the rate of increase in $\beta$  plateaued at approximately $b_f$\textgreater{}1000. This plateauing indicates that there is a point at which further increases of $b_f$ fail to be captured by the EDPVR.

Alterations in the material parameters had similar behaviors at all pressure levels; however, $a_f$ exhibited a greater influence on the EDPVR at lower pressure ($P=10$ mmHg) as noted by the larger range of $\beta$ compared to other pressure values. In contrast, $b_f$ had the greatest impact on the EDPVR at higher pressures ($P=30$ mmHg).

\subsection{Altered myocardial stiffness in various cardiac geometries}
\begin{figure}[htbp]
\centerline{\includegraphics[width=0.5\textwidth]{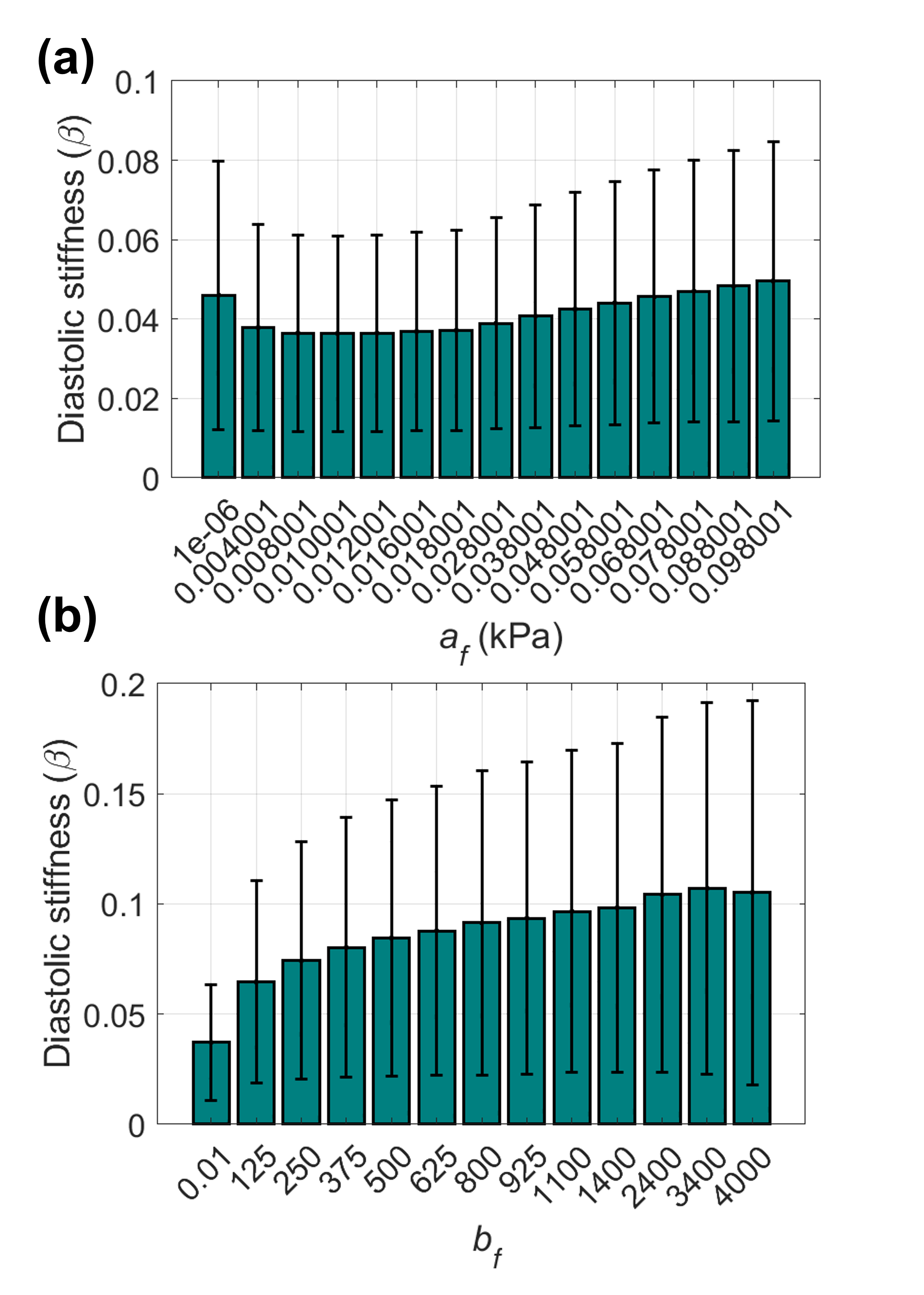}}
\caption{Exploring the correlation between EDPVR-derived stiffness metric $\beta$ and material model parameters across various geometries (n=13), in terms of mean and $\pm$standard deviation for (a) $a_f$ ($b_f$=10) and (b) $b_f$ ($a_f$=0.05kPa).}
\label{fig:geom}
\end{figure}

We extended our analysis to consider diverse geometries to further explore the relationship between effective myocardial stiffness and the EDPVR-derived metric $\beta$. The addition of various geometries indeed reduces the impact a single geometry may have on the previously presented trends. The results indicated a reduction in $\beta$ as $a_f$ increased initially, followed by inflection and increase beyond $a_f$ = 0.012 (Fig. \ref{fig:geom}a). EDPVR-derived stiffness took a sharp upward trend as $b_f$ increased and then plateaued (Fig. \ref{fig:geom}b). Importantly, these trends remained consistent with the representative observations in Figure \ref{fig:sensitivity}.

\section{Discussion}
We have presented a modeling platform that integrated subject-specific imaging data to develop high-fidelity computational cardiac models, employing a material model characterized by parameters $a_f$ and $b_f$. This work analyzed the intricate interplay between these material parameters, which control the effective myocardial stiffness, and the EDPVR-derived myocardial stiffness metric ($\beta$). To our knowledge, our study represents the first attempt to systematically elucidate the relationship between material model parameters and EDPVR-derived stiffness metrics. This exploration is critical for advancing the accuracy of diastolic stiffness metrics derived from in-vivo data. Indeed, our results indicate the need for improved stiffness metrics that take into account the complex structure and behavior of the myocardium.

\subsection{EDPVR: A gross measure of LV compliance}
Our examination of the alterations of the EDPVR in response to changes in myocardium stiffness (characterized by the material parameters) challenges the traditional use of EDPVR as an adequate measure of LV stiffness. The trends observed in our analysis highlight an additional layer of complexity regarding the capability of the EDPVR to capture the complex alterations in anisotropic myocardial mechanical behavior. Notably, the initial reduction in $\beta$ associated with increased $a_f$ and the distinct trends of $\beta$ observed at higher values of $a_f$ and $b_f$ necessitate a comprehensive reevaluation of the current understanding of the relationship between EDPVR and actual myocardial stiffness. As diastolic stiffness is a crucial metric of diastolic function in heart diseases such as diastolic heart failure and cardiomyopathy, such advancement is crucial to developing improved metrics of diastolic behavior.

\subsection{Structurally-informed measure of diastolic stiffness}


Our results highlight a need for an improved metric of diastolic myocardial stiffness. A comprehensive measure of tissue stiffness should consider cardiac geometry, myocardial tissue architecture, and subject-specific loading conditions. Indeed, the parameters we used in this study to characterize stiffness, $a_f$ and $b_f$, will be suitable metrics that accurately describe the myocardium passive behavior. Extended constitutive models may even allow for the characterization of anisotropic and regional stiffening, which is critical to understanding cardiac function in heart diseases such as myocardial infarction. While subject-specific computational models, such as those presented in this work, have been used to estimate parameters that characterize anisotropic stiffening via inverse modeling, the development of such FE models is costly and not feasible for use in clinical settings. However, recent studies have presented the possibility of using machine learning methods to predict such material parameters from imaging data commonly available in the clinical setting \cite{Babaei}. Our results highlight the need for the further development of tools that may enhance the available gross measures of LV stiffness.

\subsection{Limitations and future directions}
While our study employed 3D geometry from imaging, a limited set of parameters were utilized, including a phenomenological material model and fixed fiber architecture \cite{Usman}. Our future work will conduct sensitivity analyses concerning stiffness with varying fiber orientations and explore various material models that incorporate the heterogeneity of material properties in the myocardium.

\section{Conclusion}
We investigated the relationship between intrinsic myocardial stiffness and EDPVR-derived LV stiffness. While EDPVR-derived metrics are commonly used to evaluate the diastolic function of the LV, our results indicate non-monotonic relationships between the diastolic stiffness $\beta$ and the material properties $a_f$ and $b_f$. Image-based, in-silico-derived measures metrics hold promise in providing a more accurate assessment of myocardial behavior, and further development of tools and processes capable of bringing such metrics to the clinical setting is warranted.

\vspace{12pt}
\color{red}


\begin{thebibliography}{00}
\bibitem{Numata} Numata, Genri, et al. "A pacing-controlled protocol for frequency-diastolic relations distinguishes diastolic dysfunction specific to a mouse HFpEF model." American Journal of Physiology-Heart and Circulatory Physiology 323.3 (2022): H523-H527.
\bibitem{Bermejo2013}{Bermejo, Javier, et al. ``Diastolic chamber properties of the left ventricle assessed by global fitting of pressure-volume data: improving the gold standard of diastolic function.'' Journal of Applied Physiology (1985). 2013; 115(4): 556-68.}
\bibitem{Mehdi} Mehdi, Rana Raza, et al. "Comparison of three machine learning methods to estimate myocardial stiffness." Reduced Order Models for the Biomechanics of Living Organs. Academic Press, 2023. 363-382.
\bibitem{Babaei} Babaei, Hamed, et al. "A machine learning model to estimate myocardial stiffness from EDPVR." Scientific Reports 12.1 (2022): 5433.
\bibitem{Avaz2019}{Avazmohammadi, Reza, et al. ``A Computational Cardiac Model for the Adaptation to Pulmonary Arterial Hypertension in the Rat.'' Annals of Biomedical Engineering, 47, 138–153, 2019.}
\bibitem{Mendiola}{Mendiola, Emilio A., et al. "Right ventricular architectural remodeling and functional adaptation in pulmonary hypertension." Circulation: Heart Failure 16.2 (2023): e009768.}
\bibitem{Avazmohammadi} Avazmohammadi, Reza, et al. "A contemporary look at biomechanical models of myocardium." Annual review of biomedical engineering 21 (2019): 417-442.
\bibitem{Holzapfel} Holzapfel, Gerhard A., and Ray W. Ogden. "Constitutive modelling of passive myocardium: a structurally based framework for material characterization." Philosophical Transactions of the Royal Society A: Mathematical, Physical and Engineering Sciences 367.1902 (2009): 3445-3475.
\bibitem{Jang} Jang, Sae, et al. "Biomechanical and hemodynamic measures of right ventricular diastolic function: Translating tissue biomechanics to clinical relevance." Journal of the American Heart Association 6.9 (2017): e006084.
\bibitem{Usman} Usman, Muhammad, et al. "On the possibility of estimating myocardial fiber architecture from cardiac strains." International Conference on Functional Imaging and Modeling of the Heart. Cham: Springer Nature Switzerland, 2023.



\end{thebibliography}
\end{document}